\documentclass[letterpaper]{article} 
\usepackage{aaai23}  
\usepackage{times}  
\usepackage{helvet}  
\usepackage{courier}  
\usepackage[hyphens]{url}  
\usepackage{graphicx} 
\usepackage{natbib}  
\usepackage{caption} 
\usepackage{xspace}

\usepackage{wrapfig}
\usepackage[para,online,flushleft]{threeparttable}

\usepackage[utf8]{inputenc} 
\usepackage[T1]{fontenc}    
\usepackage{url}            
\usepackage{booktabs}       
\usepackage{amsfonts}       
\usepackage{amsmath, amsthm, amscd, bm}
\usepackage{multirow}
\usepackage{nicefrac}       
\usepackage{microtype}      
\usepackage{xcolor}         
\usepackage{adjustbox}
\usepackage{algorithm}
\usepackage{algorithmic}
\usepackage{makecell}
\usepackage{xspace}
\usepackage{tcolorbox}
\usepackage{enumitem}
\newtheorem{proposition}{Proposition}
\usepackage{caption}

\usepackage{newfloat}
\usepackage{listings}
\DeclareCaptionStyle{ruled}{labelfont=normalfont,labelsep=colon,strut=off} 
\floatstyle{ruled}
\newfloat{listing}{tb}{lst}{}
\floatname{listing}{Listing}
\title{Feature-Space Bayesian Adversarial Learning Improved Malware Detector Robustness}

\author{Bao Gia Doan\textsuperscript{\rm 1}, Shuiqiao Yang\textsuperscript{\rm 2}, Paul Montague\textsuperscript{\rm 4}, Olivier De Vel\textsuperscript{\rm 3}, Tamas Abraham\textsuperscript{\rm 4}, Seyit Camtepe\textsuperscript{\rm 3}, Salil S. Kanhere\textsuperscript{\rm 2}, Ehsan Abbasnejad\textsuperscript{\rm 1}, Damith C. Ranasinghe\textsuperscript{\rm 1}}

\affiliations{
    \textsuperscript{\rm 1}The University of Adelaide, Australia\\
    \textsuperscript{\rm 2}The University of New South Wales, Australia\\
    \textsuperscript{\rm 3}Data61, CSIRO, Australia\\
    \textsuperscript{\rm 4}Defence Science and Technology Group, Australia\\
    \{giabao.doan, ehsan.abbasnejad, damith.ranasinghe\}@adelaide.edu.au, \{shuiqiao.yang, salil.kanhere\}@unsw.edu.au, seyit.camtepe@data61.csiro.au, olivierdevel@yahoo.com.au, \{paul.montague, tamas.abraham\}@defence.gov.au
}

\begin{document}

\maketitle

\newcommand{\bao}[1]{\textcolor{magenta}{#1}}
\newcommand{\todo}[1]{\textcolor{orange}{\textbf{TODO}: #1}}
\newcommand{\heading}[1]{\noindent\textbf{#1\xspace}}

\newcommand{\rqq}[1]{
\begin{center}
\begin{tcolorbox}[width=\columnwidth, colback=white!15,left=1pt,right=1pt,top=1pt,bottom=1pt,arc=5pt,auto outer arc]
\textit{#1}
\end{tcolorbox}
\end{center}
}

\newcommand\scalemath[2]{\scalebox{#1}{\mbox{\ensuremath{\displaystyle #2}}}}

\newcommand{\feat}{\boldsymbol\phi}
\newcommand{\D}{\mathcal{D}}
\newcommand{\Z}{\mathcal{Z}}

\newcommand{\X}{\mathcal{X}}
\newcommand{\Y}{\mathcal{Y}}
\newcommand{\Da}{\mathcal{D}_{\text{adv}}}
\newcommand{\Dtilde}{\tilde{\mathcal{D}}}
\newcommand{\btheta}{\boldsymbol\theta}
\newcommand{\bTheta}{\boldsymbol\Theta}
\newcommand{\bomega}{\boldsymbol\omega}
\newcommand{\hatff}{\hat{\boldsymbol{\phi}}}

\newcommand{\tocheck}[1]{\textcolor{red}{check: #1}}
\newcommand{\damith}[1]{\textcolor{blue}{DR: #1}}
\newcommand{\ehsan}[1]{\textcolor{orange}{Ehsan: #1}}
\newcommand{\dtoprule}{\specialrule{1pt}{0pt}{\belowrulesep}
            %
            }
\newcommand{\dbottomrule}{
            \specialrule{1pt}{0pt}{\belowrulesep}%
            }

\newcommand{\bx}{\textbf{x}}
\newcommand{\bz}{\textbf{z}}
\newcommand{\bT}{\mathbf{T}}
\newcommand{\bxtidle}{\tilde{\textbf{x}}}
\newcommand{\bdelta}{\boldsymbol{\delta}}
\def\bxa{\bx_{\text{adv}}}
\def\feata{\feat_{\text{adv}}}

\def\II{\mathbb{I}}
\def\ie{\textit{i.e.}\xspace}
\def\ours{Adv-MalBayes\xspace}
\def\eg{\textit{e.g.}\xspace}
\def\etal{\textit{et al.}\xspace}

\begin{abstract}

We present a new algorithm to train a robust malware detector. Malware is a prolific problem and malware detectors are a front-line defense. Modern detectors rely on machine learning algorithms. Now, the adversarial objective is to devise alterations to the malware code to decrease the chance of being detected whilst \textit{preserving the functionality and realism of the malware}. Adversarial learning is effective in improving robustness but generating functional and realistic adversarial malware samples is non-trivial. Because: i)~in contrast to tasks capable of using gradient-based feedback, adversarial learning in a domain without a \textit{differentiable  mapping function} from the \textit{problem space} (malware code inputs) to the \textit{feature space} is hard; and ii)~it is difficult to ensure the adversarial malware is realistic and functional. 
This presents a challenge for developing scalable adversarial machine learning algorithms for large datasets at a production or commercial scale to realize robust malware detectors. We propose an alternative; perform adversarial learning in the \textit{feature space} in contrast to the problem space. We \textit{prove} the projection of perturbed, yet valid malware, in the problem space into feature space will always be a subset of adversarials generated in the feature space. Hence, by generating a robust network against feature-space adversarial examples, we inherently achieve robustness against problem-space adversarial examples. We formulate a Bayesian adversarial learning objective that captures the distribution of models for improved robustness. 
To \textit{explain} the robustness of the Bayesian adversarial learning algorithm, we \textit{prove} that our learning method bounds the difference between the adversarial risk and empirical risk and improves robustness. We show that Bayesian neural networks (BNNs) achieve state-of-the-art results; especially in the False Positive Rate (FPR) regime. Adversarially trained BNNs achieve state-of-the-art robustness. Notably, adversarially trained BNNs are robust against stronger attacks with larger attack budgets by a margin of up to 15\% on a recent production-scale malware dataset of more than \textit{20~million samples}. Importantly, our efforts create a benchmark for future defenses in the malware domain.

\end{abstract}

\section{Introduction}
We are amidst a meteoric rise in malware incidents worldwide. Malware is responsible for significant damages, both financial---in billions of dollars~\citep{anderson2019measuring}---and human costs in  loss of life~\citep{eddy_perlroth_2020}. According to statistics from Kaspersky Lab, at the end of 2020, there were an average of 360,000 pieces of malware detected per day~\citep{kaspersky}. The battle against such large incidents of malware remains an ongoing challenge and the need for automated and effective malware detection systems is a research imperative. 

Advances in Machine Learning (ML) have led to state-of-the-art malware detectors~\citep{arp2014drebin,peng2012using,harang2020sorel,raff2018malware,anderson2018ember}. 
But, ML-based models are known to be vulnerable to \textit{adversarial examples}; here, seemingly benign inputs with small perturbations can successfully evade detectors. Although adversarial examples were shown initially in the computer vision domain~\citep{szegedy2013intriguing, goodfellowExplainingHarnessingAdversarial2015, pgd, biggio2018wild}, malware is no exception. Recent attacks have crafted adversarial examples in the malware domain---so-called \textit{adversarial malware}; now, a carefully crafted malware sample with minimal changes to malware code but still able to preserve the \textit{realism} and \textit{functionality} of the malware is able to fool ML-based malware detectors to misclassify them as benign-ware. These attacks pose an emerging threat against ML-based malware detectors~\citep{grosse2017adversarial, kolosnjaji2018adversarial, kreuk2018deceiving, suciu2019exploring, pierazzi2020intriguing, demetrio2021functionality}.

\vspace{1mm}
\noindent\textbf{Problem.~}In general, adversarial learning~\cite{athalye2018obfuscated} or training with adversarial examples is an effective method to build models robust against adversarial examples. 
However, generating adversarial malware samples for training, especially at the \textit{production scale} necessary for deployable models, is non-trivial. Because:
\begin{itemize}
    \item Generation of adversarial examples in the malware domain is confronted with the \textit{inverse feature-mapping problem} where the function mapping from the \textit{problem space} (the discrete space of software code binaries) to the feature space (vectorized features) is non-differentiable~\citep{biggio2013evasion, biggio2013security, quiring2019misleading}. Hence, \textit{fast}, gradient-driven  methods to derive useful information to craft adversarial samples in the problem space are not suitable.
    \item The need to enforce malware domain constraints,  \textit{realism}, \textit{functionality}, and \textit{maliciousness} on generated perturbations in the problem space is a difficult proposition. Thus, arbitrary changes to the malware binaries are not possible because it could drastically alter the malware in a manner to break the malicious functionality of the binaries or even make it unloadable.
\end{itemize} 

Although efforts to realize robust models on discrete spaces such as discrete image or graph data exist~\citep{lee2019tight, wang2021certified}, the problem space of malware classification is significantly more challenging due to the imposed constraints in the problem space; the \textit{realism} and \textit{functionality} as well as \textit{maliciousness} of the malware must be maintained. Unfortunately, a method to scale up adversarial training with samples in the problem space to production scale datasets, especially in the case of neural networks, does not exist.

Further, despite extensive work on adversarial ML in general, very few studies have focused on the problem in the context of malware as recently highlighted by~\citet{pierazzi2020intriguing}, and a comprehensive investigation of robust defense methods in the area remains to be conducted. 

\vspace{1mm}
\noindent\textbf{Research Questions.~}Hence, in this study, we seek to answer the following research questions (RQs):

\begin{itemize}
    \item \textbf{\textit{RQ1.~}}How can we overcome the challenging problem of adversarial learning for malware at a \textit{production scale} to realize robust malware detectors against adversarial malware samples?
    
    \item \textbf{\textit{RQ2.~}}How can we formulate an adversarial learning problem for building robust malware detectors and how can we \textit{explain} the robustness and \textit{benefits}? 
    
    \item \textbf{\textit{RQ3.~}}How robust are adversarially trained malware detectors, especially against problem-space (functional, realistic and malicious) adversarial malware samples? 
\end{itemize}

\noindent\textbf{Our Approach.~}We argue that a defender is not confronted with the problems we mentioned. Because, we show that constraining the adversarial examples in the problem space to \textit{preserve malware realism, functionality and maliciousness can be turned to an advantage for defenders. The constraints make the perturbed malware in the problem space a subset of the adversarial examples in the feature space. Therefore, designing a robust method against feature-space adversarial examples will inherently be robust against constrained problem-space adversarial examples encapsulating the threats from adversarial malware.}

To construct a formulation to improve the robustness against feature-space adversarial malware examples, and ultimately problem space malware, we propose a Bayesian formulation for adversarially training a neural network: i)~with the capability to capture the distribution of models to improve robustness~\citep{Liu2016, advbnn, BAL, wicker21, Carbone21, doan22a}; 
and ii)~prove our proposed method of diversified Bayesian neural networks hardened with adversarial training
bounds the difference between the adversarial risk and the conventional empirical risk to \textit{theoretically explain} the improved robustness. 

Moreover, just recently, security researchers with domain expertise placed significant effort into providing features\footnote{Notably, a negligible computation time of 160~ms, on average, is required to derive vectorized features as described in the \textbf{Appendix}.} for malware samples at a production scale of more than 20 million samples~\citep{harang2020sorel, anderson2018ember}--the \textsf{SOREL-20M} dataset. However, \textit{the robustness of networks built on these extracted features in the face of evasion attacks are yet to be understood}. Therefore, our study to investigate production scale adversarial learning is timely and we focus our efforts to investigate methods using the \textsf{SOREL-20M} dataset.

\vspace{1mm}
\noindent\textbf{Our Contributions.~}
{To address the problem of building robust malware detectors, we make the following contributions:}
\begin{enumerate}[itemsep=0em]
    \item We \textit{prove} the projection of perturbed yet, valid malware, in the problem space (the discrete space of software code binaries) into the feature space will be a subset of feature-space adversarial examples. Thus, a robust network against feature-space attacks is inherently robust against problem-space attacks. Our work provides \textit{a theoretically justified basis for adversarially training malware detectors in the feature space}. Further, to corroborate our proof, we empirically demonstrate networks trained on feature-space adversarials are robust against functional and realistic problem-space adversarial malware (\textbf{\textit{RQ1}}).
    
    \item Hence, to improve robustness in the \textit{problem space} we propose performing adversarial learning in the \textit{feature space} and formulate a Bayesian Neural Network (BNN) adversarial learning objective that captures the distribution of models for improved robustness. The algorithm is capable of learning from production scale feature-space datasets of up to \textit{20 million samples} (\textbf{\textit{RQ1}} and \textbf{\textit{RQ2}}).  
    
    \item We also \textit{prove} hardening BNNs with adversarial examples bounds the difference between the adversarial risk and the empirical risk to explain the improved robustness (\textbf{\textit{RQ2}}). 
    
    \item We empirically demonstrate Bayesian Neural Networks capturing model diversity to improve the performance of malware classifiers and adversarially trained BNNs to generate more robust models against the threat of adversarial malware. Adversarially trained BNNs achieve new benchmarks for state-of-the-art robustness---especially against unseen, stronger, attack samples (\textbf{\textit{RQ3}}). 
\end{enumerate}

\noindent\textbf{Scope.~}Notably, in our study, we focus on Windows Portable Executable (PE) malware for two reasons: i)~Windows is the most popular operating system for end-users worldwide,  and PE-file malware is the earliest and most studied threat in the wild~\citep{schultz2000data}, making a robust method to detect adversarial PE files a significant contribution to security research; and ii)~the intuition and methodology behind Windows PE malware can be applied and transferred to other file formats and operating systems, such as PDF malware or malware for Linux and Android systems {(see the \textbf{Appendix})}.

\section{Background and Related Work}

\noindent\textbf{Machine Learning Methods in the Malware Domain.~}Malware detection is moving away from hand-crafted approaches relying on rules toward machine learning (ML) techniques~\citep{schultz2000data,saxe2015deep, raff2018malware, krvcal2018deep}.
Recently, 
MalConv~\citep{raff2018malware} adopted a Convolutional Neural Network (CNN) based architecture design with a learnable, but  non-differentiable, embedding space for malware detection from raw byte sequences. The adoption of a CNN for malware detection was also proposed in~\citep{krvcal2018deep}. However, training malware detectors on raw byte sequences (arbitrary number, often millions, of bytes) is computationally expensive and time-consuming. In addition, as we discussed earlier, it is non-trivial to craft realistic adversarial examples on raw byte sequences to realize a robust network on large-scale datasets. Consequently, recent work has employed problem space to feature space mapping functions together with  feed-forward neural networks to build benchmark models for the large-scale \textsf{SOREL-20M} dataset~\cite{harang2020sorel}. 

LUNA~\citep{backes17} proposed a simple linear Bayesian model for an Android malware detector, which preserves the concept of uncertainty, and shows that it helps to reduce incorrect decisions as well as improve the accuracy of classification. The benefit of a Bayesian classifier is to handle ML tasks from a stochastic perspective, where all weight values of the network are probability distributions. More recently, \citet{nguyenLeveragingUncertaintyImproved2021} investigated the application of uncertainty and Bayesian treatment to improve the performance of malware detectors on neural networks.   

\vspace{1mm}
\noindent\textbf{Adversarial Malware (Adversarial Examples in the Malware Domain).~}ML-based classifiers are shown to suffer from \textit{evasion} attacks, via \textit{adversarial examples}~\citep{goodfellowExplainingHarnessingAdversarial2015}. 
Recently, adversarial examples were demonstrated in the \textit{problem space}~\citep{grosse2016adversarial, xu2016automatically, grosse2017adversarial,hu2017generating,kolosnjaji2018adversarial,kreuk2018deceiving,suciu2019exploring}. 
In particular, \citet{kolosnjaji2018adversarial} proposed a method to append bytes to the end of the binary PE file, while~\citet{kreuk2018deceiving} exploited the regions within the executable which are not mapped to memory to construct adversarial malware. These methods intend to make modifications that do not affect the intended behavior of the executable.
Suciu \etal~\citep{suciu2019exploring} \textit{adopted} FGSM~\citep{goodfellowExplainingHarnessingAdversarial2015} to show the generalization properties and effectiveness of adversarial examples against a CNN-based malware detector, MalConv, trained with small-scale datasets. \citet{suciu2019exploring} highlighted the threat from adversarial examples as an alternative to evasion techniques such as runtime packing, but showed that models trained on small-scale datasets did not generalize to robust models; hence, \textit{emphasizing the importance of training networks on production-scale datasets}.

\vspace{1mm}
\noindent\textbf{Improving Model Robustness.~}Among methods for improving the robustness of models~\citep{pgd, chen2020training, fischer2019dl2}, adversarial training~\citep{pgd} and 
its variants 
are shown to be one of the most effective and popular methods to defend against adversarial examples~\citep{athalye2018obfuscated}. The goal of adversarial training is to incorporate the adversarial search within the training process and, thus, realize robustness against adversarial examples at test time. 
In particular, recently, Bayesian adversarial learning has been investigated and adopted in the computer vision domain to propose to improve the robustness of models against adversarial examples~\citep{Liu2016, BAL, advbnn, wicker21, Carbone21, doan22a}. 

Adversarial learning was explored in the malware domain in~\citep{al2018adversarial} to generate a robust detector for binary encoded malware. However, the computational cost to realize realistic, adversarial raw byte representations is prohibitively expensive~\citep{suciu2019exploring, pierazzi2020intriguing} for adversarial learning. 

\vspace{1mm}
\noindent\textbf{Summary.~}We recognize that: i)~a method capable of scaling up the adversarial training of neural networks in the problem space to production scale datasets does not exist; ii)~a Bayesian adversarial learning objective that captures the distribution of models could provide improved robustness; however iii)~such a formulation requires overcoming the challenging problem of generating problem-space adversarial examples at production scales. 

In what follows, we begin with a problem definition, a theoretical basis for employing feature-space adversarial learning as an alternative to problem-space, followed by the formulation of a Bayesian adversarial learning objective and experimental results validating our claims and demonstrating state-of-the-art performance and robustness.

\section{Problem Definition}

\textbf{Threat model.~}We assume an attacker with \textit{perfect knowledge} (white-box attacker)~\citep{biggio2013security}, in which the attacker knows \textit{all} parameters including feature set, learning algorithm, loss function, model parameters/hyperparameters, and training data. The reason for considering the strongest, perfect-knowledge adversary is because, even if access to the model is not possible, or the model is not publicly available, an adversary can employ a reverse engineering approach such as~\citep{tramer2016stealing, rolnick20a, carlini2020cryptanalytic} to extract the model. And, defending against such attacks is challenging. 
The attacker's objective is to \textit{evade detection}. Their capability is to modify the features at test time. 

\heading{Problem-Space Attacks}. 
We consider the \textbf{problem space} $\mathcal{Z}$ which refers to the \textit{input space} of real objects of a considered domain such as software code binaries.
First $\Z$ must be transformed into a compatible format such as numerical vector data~\citep{anderson2018ember, harang2020sorel} for ML to process. Then, a \textbf{feature mapping} is a function $\Phi: \mathcal{Z} \rightarrow \mathcal{X} \subseteq \mathbb{R}^n$ that maps a given problem-space software code binary $\bz \in \Z$ to an n-dimensional feature vector $\bx \in \X$ in the \textbf{feature space} such that $\Phi(\bz) = \bx$.

Normally, attackers have to apply a transformation on $\bz$ to generate $\bz'$ such that $\Phi(\bz')$ is very close to $\bx'$ in the feature space. Formally, given a problem-space object $\bz \in \Z$ with label $y \in \Y$, the goal of the adversary is to find the \textit{transformation} $\bT: \Z \rightarrow \Z$ (\eg addition, removal, modification) such that $\bz' = \bT(\bz)$ is classified as a class $t \neq y$. In the malware domain, the adversary has to search in the problem space that approximately follows the gradient in the feature space. However, this is a major challenge that complicates the application of gradient-driven methods to the problem-space attacks--- so-called \textbf{inverse feature-mapping problem}~\citep{quiring2019misleading, biggio2013evasion, pierazzi2020intriguing} where the function $\Phi$ in the software domain---our focus---is typically not invertible and not differentiable, \ie there is no one-to-one mapping from the adversarial examples in the feature space $\bx + \bdelta$ to the corresponding adversarial problem-space object $\bz'$. In addition,  the generated object $\bT(\bz)$ must be realistic and valid~\citep{suciu2019exploring}.
Thus, the search for adversarial examples in the problem space (software) cannot be a purely gradient-based method, hindering the adoption of well-known adversarial attacks in other domains such as computer vision. 
To achieve a realistic adversarial objective, the search for adversarial examples in the problem space has to be constrained in \textbf{problem-space constraints} denoted by $\Omega$. We remark that the constraints on the problem space are well defined and can be found in~\citep{biggio2018wild, quiring2019misleading, xu2016automatically, pierazzi2020intriguing}, we mentioned here, for completeness, that there are at least four main types of problem-space constraints including Preserved semantics, Plausibility, Robustness to Processing and Available Transformation explained in detail by~\citet{pierazzi2020intriguing}.

\vspace{1mm}
\heading{Feature-Space Attacks.} To alleviate the problems with problem space attacks, we propose an alternative that uses feature space. We note that all definitions of feature-space attacks are well defined and consolidated in related work~\citep{biggio2018wild, carlini2017towards, grosse2017adversarial}. In this paper, we use a popular feature mapping function provided in the EMBER dataset~\citep{anderson2018ember} to map raw bytes of software to a vector of $n=2381$ features. 
A \textbf{feature-space attack} is then to modify a feature-space object $\bx \in \X$ to become another object $\bx' = \bx + \bdelta$ where $\bdelta$ is the added perturbation crafted with an \textit{attack objective function} to  miclassify $\bx'$ into another class $t \neq y$ where $y \in \Y$ is the ground-truth label of $\bx$. We note that in the malware domain (a binary classification task), the intuition of the attackers is to make the malware be recognized as benign ware. These modifications has to follow \textbf{feature-space constraints}.
We denote the constraints on feature-space modifications by $\Upsilon$.
Given a sample $\bx \in \X$, the feature-space modification, or perturbation $\bdelta$ must satisfy $\Upsilon$. This constraint $\Upsilon$ reflects the realistic requirements of problem-space objects.
In the malware domain, feature perturbations $\bdelta$ can be constrained $\bdelta_{lb} \leq \bdelta \leq \bdelta_{ub}$~\citep{pierazzi2020intriguing}.

\begin{figure}[ht]
    \centering
    \includegraphics[width=.5\linewidth]{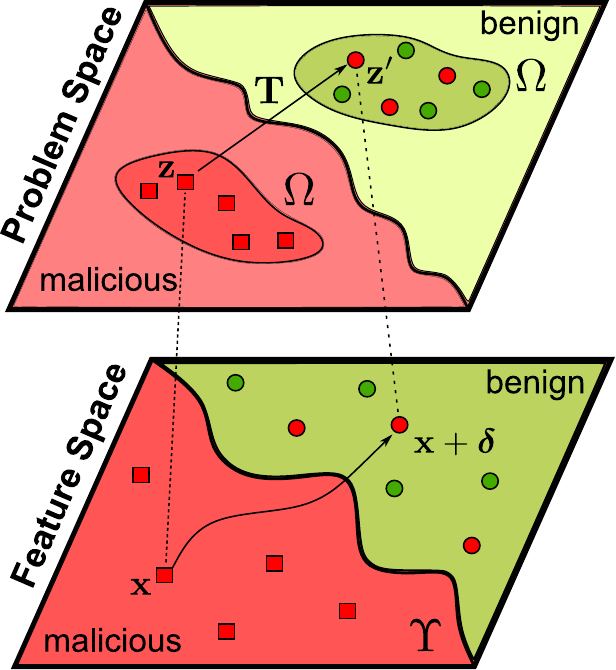}
    \caption{Illustrative example of adversarial examples. The adversarial example $\bx + \bdelta$ is derived from $\bx$ in the feature space and its projection to problem-space constraints (which is more restrictive) determined by $\Omega$ is $\bz'$. The color in the background illustrates the decision regions where red color is for malware and green is for benign programs. The solid arrow in Feature Space represents the gradient-based attack to transform a malware $\bx$ to $\bx + \bdelta$, projected to the problem-space constraints as $\bz'$ to be misdetected as a benign program.}
    \vspace{-5mm}
    \label{fig:fig1}
\end{figure}

\section{Theoretical Basis For Feature-Space Adversarial Learning}

We highlight that the realistic assumption of problem-space attacks makes the constraints imposed by $\Omega$ stricter or equal to those imposed by $\Upsilon$ (illustrated in Figure~\ref{fig:fig1}). Following the necessary condition for problem-space adversarial examples as stated in~\citet{pierazzi2020intriguing}, we have:

\noindent\textbf{\textit{Lemma 1.}} If there exists an adversarial example in the problem space ($\bz'$) that satisfies the constraints $\Omega$, then there will be a corresponding adversarial example in the feature space ($\bx'$) under the constraints $\Upsilon$. More formally, by abusing notation from model theory to use
$\models$ to indicate an instance ``satisfies'' constraints, and write $\bz' \models \Omega$ and $\bx' \models \Upsilon$, we have:
 \begin{align*}
\exists \bz': \bz' \models \Omega, \quad p\big(y \mid \Phi(\bz'), \btheta\big) = p(y\mid\Phi(\bT(\bz)), \btheta),  \\ \quad \quad p(y\mid\Phi(\bT(\bz)), \btheta) < 0.5 \\
\Rightarrow \exists \bx'= \bx + \bdelta: \bx' \models \Upsilon, \quad p(y\mid \bx',\btheta) < 0.5
\end{align*}
where $\bT$ is the transformation in the problem
space to craft adversarial examples, 
$p(y\mid \bx, \btheta)=\text{sigmoid}(f(\bx;\btheta))$
is the output of a
\texttt{sigmoid} function
applied to the output of the neural networks $f$ parameterized by $\btheta$, $p(y\mid \bx, \btheta)=0.5$ is the threshold for malware detection where the predicting $p(y\mid \bx, \btheta)=0$ is recognized as benign whilst $p(y\mid \bx, \btheta)=1$ indicates a malware,
$\Omega, \Upsilon$ are, respectively, the problem-space and feature-space constraints, and $\Phi(\cdot)$ is the function that maps the problem space to feature space.

The proof of Lemma 1 is in \textbf{Appendix}~\ref{appd:proof_theorem1}.
From \textbf{Lemma 1}, if there exists an attack in the problem space, then there exists a corresponding attack in the feature space. By contraposition, if there does not exist an attack in the feature space, there does not exist an attack in the problem space. However, we know that the opposite is not true: if there does not exist an attack in the problem space (e.g. due to functionality), there still exists an attack in the feature space. Thus, we can derive:

\vspace{1mm}
\noindent\textbf{\textit{Corollary 1}}. The adversarial examples generated from constrained problem-space adversarial examples (imposed by $\Omega$) are in a subset of feature-space adversarial examples (imposed by $\Upsilon$).

\vspace{1mm}
\noindent\textbf{\textit{Corollary 2}}. Detectors robust against feature-space adversarial examples (imposed by $\Upsilon$) are robust against constrained problem-space adversarial examples (imposed by $\Omega$). 

Built upon these Corollaries, we propose to find a learning method robust against \textit{feature-space adversarial malware}. On the one hand, adversarial training~\citep{pgd} and its variants are shown to be one of the most effective and popular methods to defend against adversarial examples~\cite{athalye2018obfuscated}. 
On the other hand, Bayesian neural networks~\citep{mackay1992practical, ritter2018scalable, izmailov2021bayesian} with distributions placed over their weights and biases enabling the principled quantification of the uncertainty of their predictions are shown to be a robust method against adversarial examples.
Thus, in this paper, demonstrating that robustness against feature-space adversarial examples is inherently robust against problem-space real malware. We propose to incorporate adversarial training with Bayesian neural networks to seek the first principled method of Bayesian adversarial learning to realize a robust malware detector without the difficulties of inverse feature-mapping and preserving semantics and functionalities of real malware samples. We name our method \ours, and the method is efficient enough to be scaled up to a large production scale of adversarial training data of 20 million adversarial samples with the pre-extracted feature set of SOREL-20M dataset~\citep{harang2020sorel}.

\section{Bayesian Formulation for Adversarial Learning}
\label{sec:bayesian_formulation}

The goal of Bayesian adversarial learning is to find the posterior distribution using Bayes theorem: 
\begin{equation*}
p(\btheta\mid\Da)={\prod_{(\bxa,y)\sim\Da}p(y\mid\bxa,\btheta)p(\btheta)}/{Z}
\end{equation*}
where $Z$ is the normalizer, $\Da$ is the adversarial dataset obtained by generating adversarial examples from the benign dataset $D$ using adversarial generation such as Eq.~\eqref{eq:a_pgd}.

We consider $p(y\mid\bxa,\btheta)=\text{sigmoid}(f(\bxa;\btheta))$ to produce a binary prediction in malware detection.
Notably, Eq.~\eqref{eq:a_pgd} is the Expectation-over-Transformation (EoT) PGD attack~\citep{athalye2018synthesizing, rol2019comment}, which is slightly different from the usual PGD attack~\citep{pgd}. As has been highlighted in~\citet{rol2019comment}, the EoT attack is better able to estimate the gradient of the stochastic Bayesian models:
{\small
\begin{align}
    \label{eq:a_pgd}
    \scalemath{0.95}{\bx^{t+1}=\Pi_{\varepsilon_{\max}}\left\{\bx^{t}+\alpha \cdot \operatorname{sign}\left({\mathbb{E}_{\btheta}}\left[\nabla_{\bx} \ell\left(f\left(\bx^{t} ; \btheta\right), y_{o}\right)\right]\right)\right\}.}
\end{align}
}%
where $\varepsilon_{\max}$ is the maximum attack budget, $\Pi_{\varepsilon_{\max}}$ is the projection to the set $\left\{\bx \mid\left\|\bx-\bx_{o}\right\|_{\infty} \leq \varepsilon_{\max}\right\}$,  $\ell$ is the loss function (typically cross-entropy), $f$ is the neural network, $\mathbf{x}$ is the input, $\btheta$ is the network parameter, and $y$ is the ground-truth label. 
In this attack, an attacker starts from $\bx^0=\bx_{o}$ and conducts projected gradient descent iteratively to update the adversarial example.

However, as highlighted in~\citet{izmailov2021bayesian}, the posterior over a Bayesian neural network is extremely high-dimensional, non-convex and intractable. Thus, we need to resort to approximations to find the posterior distribution. In this work, we propose using Stein Variational Gradient Descent (SVGD)~\citep{Liu2016} for two reasons. First, this approach learns multiple \emph{network parameter particles} in parallel for faster convergence. Second, there is a \emph{repulsive factor} in the method to encourage the diversity of parameter particles that helps to prevent mode collapse --- a challenge of posterior approximation. To further demonstrate the robustness of our chosen Bayesian method, we compare \ours with previous BNNs~\citep{advbnn} in the \textbf{Appendix} Table~\ref{tab:compare_stochastic}.

We consider $n$ samples from the posterior (\ie parameter particles). The variational bound is minimized when gradient descent is modified as:
 \begin{multline*}
 \label{eq:grad_update}
     \btheta_i  =   \btheta_i - \epsilon_i \hatff{}^*(\btheta_i)  \\
\text{\quad with} \quad
\hatff{}^*(\btheta) = \sum_{j=1}^n\big[  k(\btheta_j, \btheta)  \nabla_{\btheta_j} \ell(f(\bxa;\btheta_j),y)  \,\,\,\\ -\frac{\gamma}{n}\nabla_{\btheta_j} k(\btheta_j, \btheta)\big]\,.
\end{multline*}
 Here, $\btheta_i$ is the $i$th particle, $k(\cdot, \cdot)$ is a kernel function that measures the similarity between particles and $\gamma$ is a hyper-parameter.
 The parameter particles are encouraged to be dissimilar to capture more diverse samples from the posterior thanks to the kernel function. This is controlled by a hyper-parameter $\gamma$ to manage the trade-off between diversity and loss minimization. Following~\citep{Liu2016}, we use the RBF kernel $\scalemath{.9}{k(\btheta, \btheta')=\exp \left(-{\left\|\btheta-\btheta^{\prime}\right\|^{2}}/{2 h^{2}}\right)}$ and take the bandwidth $h$ to be the median of the pairwise distances of the set of parameter particles at each training iteration. 

At the inference stage, given the test data point $\bx^*$, we can get the prediction by approximating the posterior using the Monte Carlo samples as:
{\small\begin{align*}
    p ( y^* \mid & \mathbf{x}^*, \Da ) = \int p(y^* \mid \mathbf{x}^*, \boldsymbol{\theta}) p(\boldsymbol{\theta} \mid \Da) d \boldsymbol{\theta} \\
    & \approx \frac{1}{n} \sum_{i=1}^{n} p(y^* \mid \mathbf{x}, \boldsymbol{\theta}_{i}), \quad \boldsymbol{\theta}_{i} \sim p(\boldsymbol{\theta} \mid \Da)\,,
\end{align*}}
where $\btheta_i$ is an individual parameter particle. 
Notably, we acknowledge that it is critical to have diverse parameter particles. Averaging over diverse and uncorrelated predictors was shown to improve network performance~\citep{jacobs1991adaptive,wolpert1992stacked, breiman1996bagging}. In the adversarial setting, 
when integrating out the parameters in our Bayesian formulation, we implicitly remove the vulnerabilities arising from a single choice of parameter existing in traditional neural networks, and hence improve the robustness.

\subsection{Adversarial Risk is Bounded with the Bayesian Formulation}

In this section, to explain the robustness of the Bayesian adversarial learning method that we propose, 
we prove that training the network with the Bayesian adversarial learning method bounds the difference between the adversarial risk and the empirical risk. This is important, because, now the risk of misclassification on adversarial examples is as the same as that of benign ones; hence eliminating the vulnerability of adversarial examples and reduce the risk of misclassification of adversarial examples to the generalization ability of the classifier. Notably, improving the generalization ability of the classifier is not our focus.

In this context, we make no specific assumption on the distribution of either the adversarial examples or the perturbations, to provide a generic defense approach. The only assumption we make is that the distribution of the data and the corresponding adversarial examples are sufficiently close. This is a mild and reasonable assumption because the idea of adversarial learning is that the added perturbation does not change the perceived samples or the distribution of the samples. Thus, we consider the bound of $\left|R_{\text{adv}}-R\right|$ where the empirical risk
$R=\mathbb{E}_{\btheta}\left[\mathbb{E}_{(\bx,y)\sim\D}\left[ \mathbb{E}_{y'\sim p(y\mid\bx,\btheta)}\left[\mathbb{I}(y=y')\right]\right]\right] $    
and the adversarial risk $R_{\text{adv}}=\mathbb{E}_{\btheta}\left[\mathbb{E}_{(\bxa,y)\sim\Da}\left[ \mathbb{E}_{y'\sim p(y\mid\bxa,\btheta)}\left[\mathbb{I}(y=y')\right]\right]\right]$

\begin{proposition}\label{prop1}
The difference between the adversarial risk (denoted by $R_{\text{adv}}$) and the empirical risk (denoted by $R$) of a classifier when trained on the observed training set and its adversarial counterparts is bounded, \ie
{\small\begin{eqnarray*}
&&\left| R_{\text{adv}}-R \right| \leq \tau, \notag \\
\text{where}~~~~~\tau&=&1 -\mathbb{E}_{(\bx,y)\sim\D} \Bigg[ \exp \bigg( \mathbb{E}_{\btheta}[r_{\btheta}(\bx,\bxa,y)] 
\bigg) \Bigg], \notag \\
r_{\btheta}(\bx,\bxa,y)&=&\sum_{c}^{K}p(y=c\mid\bx,\btheta)\log(p(y=c\mid\bxa,\btheta))\,.
\end{eqnarray*}}
Here, $\bxa$ denotes the adversarial example obtained from $\bx$.
\end{proposition}
We can see that the difference between the empirical risk and the adversarial risk is minimized when the upper bound is minimized. Notably, as we know that $1-\exp(-z)$ is a monotonically increasing function, and $1-\exp(-z)\leq z$, to avoid computational instabilities and gradient saturation, we consider minimizing the upper bound without the exponential function. Thus, to minimize the upper bound, our main learning objective (in Algorithm~\ref{alg:alg1} in the \textbf{Appendix}) is to:
\begin{itemize}[label={},leftmargin=*]
    \item \textit{Minimize cross entropy for the adversarial examples}. This corresponds to matching the prediction from the adversarial data to that of the observations. Since $(\bx,y)$ is given in the training, we simply minimize the entropy of the adversarials.
\end{itemize}
\emph{Sketch of the Proof.} We simplify the difference between the risks by considering that the difference between individual mistakes is smaller than their product, \ie $$
\begin{aligned}
\mathbb{E}_{y_1\sim p(y\mid\bx,\btheta)}\left[\mathbb{E}_{y_2\sim p(y\mid\bxa,\btheta)}\left[\mathbb{I}[y\neq y_1]-\mathbb{I}[y\neq y_2]\right]\right] \\
\leq \mathbb{E}_{y'\sim p(y\mid\bxa,\btheta)}\left[\mathbb{E}_{y'\sim p(y\mid\bxa,\btheta)}\left[\mathbb{I}[y_1\neq y_2]\right]\right]\\ \leq 1-\sum_{c=1}^K p(y=c\mid\bx,\btheta)p(y=c\mid\bxa,\btheta)\,.
\end{aligned}$$ 
We then use Jensen's inequality when using $\exp(\log(\cdot))$ to obtain the upper bound. The complete proof is provided in \textbf{Appendix}~\ref{sec:appd_proof}. We empirically evaluate this difference of risk and illustrate the results in Figure~\ref{fig:R_gap} in the \textbf{Appendix}.

\section{Experiments and Results}

\noindent\textbf{Classifiers.} To validate our proposed method \ours, we conduct experiments on different neural networks. We employ the Feed Forward Neural Network (FFNN) classifier provided in the SOREL-20M dataset~\citep{harang2020sorel}. This network architecture is also used for the experiments on the EMBER dataset~\citep{anderson2018ember}. Our network implementation uses the default configuration provided in~\citep{harang2020sorel}. We also adopt the architecture of FFNN to design the Bayesian Neural Network (BNN). The details of the network architecture are in \textbf{Appendix}. Then, we harden the FFNN and BNN with adversarial examples to generate the Adv-FFNN model and \ours. In addition, we also employ baseline networks including LightGBM~\citep{anderson2018ember} and  MalConv~\citep{raff2018malware} for comparison. We compare their performance on malware datasets (no attacks) and its adversarial counterparts (adversarial malware designed to evade detectors) to evaluate the \textit{detector performance} and \textit{robustness}. The values of the attack budgets used for training and testing are detailed in  Table~\ref{table:hyperparameters} in the \textbf{Appendix}.

\vspace{1mm}
\noindent\textbf{Datasets}.
In this paper, we use the two largest publicly available corpora for malware detection, namely:
\begin{itemize}
    \item The production scale dataset Sophos AI \textbf{\textsf{SOREL-20M}}~\citep{harang2020sorel} containing 20 million pre-extracted samples.
    \item \textbf{\textsf{EMBER}}~\citep{anderson2018ember} dataset designed to be \textit{more challenging} for ML-based classifiers.
\end{itemize}
We detail these datasets in \textbf{Appendix}~\ref{appd:dataset}. 

\begin{figure}[h]
    \centering
    \includegraphics[width=\linewidth]{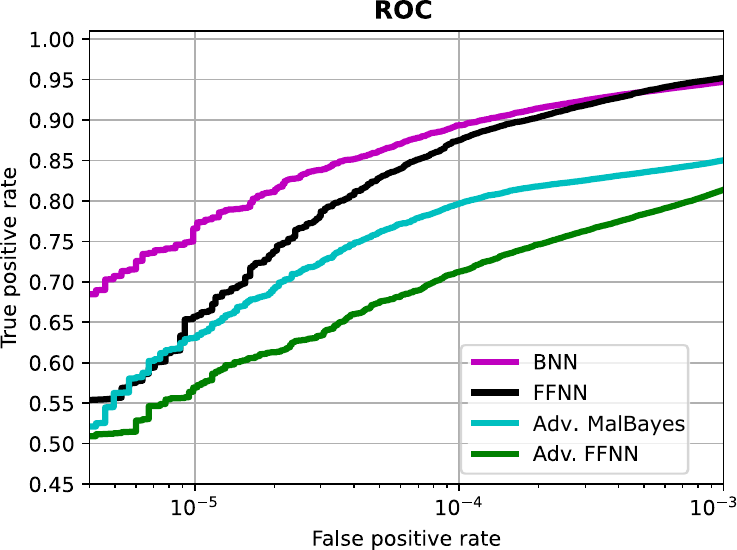}
    \caption{Performance of neural network classifiers in the absence of adversarial examples in the \textsf{SOREL-20M} dataset. 
    }
    \label{fig:benign_performance}
    \vspace{-3mm}
\end{figure}

\noindent\textbf{Results.~}We present our results by reporting: i) performance of the given classifiers on malware detection tasks (no attacks setting) using ROC (receiver operating characteristic curve); and ii)~robustness (under evasion attacks with adversarial malware). We detail these metrics in the \textbf{Appendix}.

\vspace{1mm}
\noindent\textbf{Performance (no attacks).~}Performance of the  classifiers in the absence of attacks are shown in Figure~\ref{fig:benign_performance} with additional details reported in Table~\ref{tab:noattacks} in the \textbf{Appendix}.
The ROC curves in Figure~\ref{fig:benign_performance} report the True Positive Rate (\ie the percentage of correctly-classified malware samples) as a function of the False Positive Rate (FPR, \ie the percentage of misclassified benign samples) for each classifier. From the figure, we can see that Bayesian neural networks of the same network architecture as FFNNs achieve better performance (compare BNN vs. FFNN and \ours vs. Adv-FFNN). Notably, the BNNs outperformed the FFNN counterparts with a large margin in the detection rate (of up to 20\%) under low-FPR regimes. Notably, in Table~\ref{tab:noattacks} in the \textbf{Appendix}, we also show that BNNs built on feature-space samples achieve better performance compared with the popular ML-based malware detector built on problem-space samples (MalConv)~\citep{raff2018malware} and its recently updated version in AAAI-21 (MalConv w/ GCG)~\citep{raff2020classifying}.

\vspace{1mm}
\noindent\textbf{Robustness (against Feature-Space Adversarial Examples)}. To evaluate the robustness of the investigated classifiers, we apply the PGD attack from Equation~\eqref{eq:a_pgd} on malware samples with increasing attack budgets. Results for the robustness of given classifiers under different attack budgets 
are reported in Table~\ref{tab:robustness_SOREL}. Notably, \ours outperforms the adversarially trained FFNN on both the production scale (\textsf{SOREL-20M}) and challenging (\textsf{EMBER}) datasets, especially under increasing attack budgets. This is significant because the problem with malware is that they are evolving extremely fast \eg there are hundreds of thousands of new malware samples every day~\citep{kaspersky}. Further, results in  Figure~\ref{fig:attack_results} in the \textbf{Appendix} illustrate, as expected and in line with the findings in the literature~\citep{pgd, carlini2017towards, goodfellowExplainingHarnessingAdversarial2015}, the adversarially trained networks are significantly more robust than non-adversarially trained counterparts. 

\begin{table}[h]
\centering
\resizebox{\linewidth}{!}{%
\begin{tabular}{cccccccc}
\dtoprule 
\multirow{2}{*}{Dataset} & \multirow{2}{*}{Networks} & \multicolumn{6}{c}{Attack budget} \\ \cmidrule{3-8}
 &  & \textit{0}      & \textit{0.03}  & \textit{0.05}    & \textit{0.1}   & \textit{0.2}   & \textit{0.3}   \\ \dtoprule
\multirow{2}{*}{\makecell{SOREL\\20M}} & Adv-FFNN               & 95.38  & 93.31 & 89.92 & 47.74 & 17.34 & 13.3  \\ \cmidrule{2-8}
& \ours                & \textbf{95.52}  & \textbf{94.20} & \textbf{90.53}   & \textbf{62.86} & \textbf{25.42} & \textbf{23.10} \\ \dbottomrule
\multirow{2}{*}{EMBER} & Adv-FFNN               & 86.88  & 82.44 & 79.48 & 64.00  & 51.32 & 42.77  \\ \cmidrule{2-8}
& \ours                & \textbf{89.17}  & \textbf{86.73} & \textbf{84.79}   & \textbf{78.03} & \textbf{63.06} & \textbf{52.63} \\ \dbottomrule
\end{tabular}%
}
\caption{Robustness of networks against adversarial malware generated with increasing attack budgets.}
\label{tab:robustness_SOREL}
\label{tab:robustness_EMBER}
\end{table}

\noindent\textbf{Robustness (against Problem-Space Adversarial Malware).~}In this section, we evaluate the robustness of different networks against functional, malicious and real adversarial malware in the problem space. We employ two evaluation sets. Set A includes real malware collected from a previous study~\cite{mantovani2020prevalence} and includes 7137 virus samples. We generate the real \textit{adversarial} malware samples by utilizing the constant padding attack method proposed by~\citep{Fleshman} used to win the machine learning static evasion competition~\citep{DEFCON}. In particular, 100,000 constant bytes valued \texttt{0xA9} were added to a new section of PE files to ensure the malicious functionality is not altered. The results in Table~\ref{tab:ps-ae} show that this attack can significantly degrade the performance of the popular ML-based malware detector MalConv~\citep{raff2018malware}, however the LightGBM~\citep{anderson2018ember} model is still robust against this attack (confirming the previous result obtained by~\citep{Fleshman}). 
Set B consists of the recent release by~\cite{erdemir2021adversarial}. This includes 1001 real adversarial malware samples generated using the Greedy Attack method shown to be stronger than the constant padding attack~\citep{Fleshman}. The results reported in Table~\ref{tab:ps-ae} show that 
the Greedy Attack successfully fools the LightGBM model and downgrades its robustness to 11.2\%. Notably, evaluations under both sets show the adversarially trained networks on \textit{feature-space} adversarial samples (\ie Adv-FFNN and \ours) maintained their  robustness. Importantly, \ours achieved very high \textit{robustness under both attack datasets and is a clear demonstration of the effectiveness of our approach and the validity of the theoretical basis for training with feature-space adversarial samples}.

\begin{table}[h]
\centering
\resizebox{\linewidth}{!}{
\begin{threeparttable}
\begin{tabular}{ccccccc}
\dtoprule

           & LightGBM & MalConv & FFNN  & BNN & \makecell{Adv-\\FFNN} &  \makecell{Adv-\\MalBayes} \\ \midrule
Set A & 92.5\%  & 29.2\% & 69.5\% & 72.5\% & 92.6\% &  \textbf{99.9\%}  \\ \midrule
Set B & 11.2\%  & -\tnote{1} & 74.9\% & 83.1\% & 91.8\% &  \textbf{99.9\%}  \\ \bottomrule
\end{tabular}
\begin{tablenotes}
\item[1] The released set is vectorized features, not applicable for MalConv.
\end{tablenotes}
\end{threeparttable}
}
\caption{Comparing the robustness of detectors against \textit{real} and \textit{unseen} adversarial malware (problem-space attacks). }
\label{tab:ps-ae}
\end{table}

\noindent\textbf{The Impact of Number of Parameter Particles.~} We investigate the contribution of the number of parameter particles to the robustness of the networks and report the results in Table~\ref{tab:ablative_particle} in the \textbf{Appendix}. The robustness of the BNNs is improved when more particles capable of modeling the multi-modal posterior are employed. {Thus, increasing the number of parameter particles may further improve the network's robustness.} 

\noindent\textbf{Transferability of Robustness.~}We also evaluate the robustness of the BNN trained on PGD $L_\infty$, and its transferability to other attacks, such as FGSM~\citep{goodfellowExplainingHarnessingAdversarial2015}.
Table~\ref{tab:transfer_robustness} shows that the network trained with PGD $L_\infty$ is robust to other attacks, in line with~\citep{pgd}, where PGD $L_\infty$ is considered as the `universal' attack. Consequently, we can expect our method to improve robustness against a wide range of other adversarial example generation methods~\citep{suciu2019exploring, kolosnjaji2018adversarial, kreuk2018deceiving} adopting the FGSM method to attempt to generate problem space malware samples. 

\begin{table}[h]
\vspace{-2mm}
\centering
\resizebox{\linewidth}{!}{%
\begin{tabular}{ccccccc}
\dtoprule
 \multirow{2}{*}{\ours Networks} & \multicolumn{6}{c}{{Attack budget}} \\ \cmidrule{2-7}
 & \textit{0}      & \textit{0.03}  & \textit{0.05}    & \textit{0.1}   & \textit{0.2}   & \textit{0.3}   \\ \dtoprule
PGD $L_\infty$              & 96.29  & 94.97 & 92.19   & 69.96 & 35.20 & 30.79 \\ \midrule

FGSM              & -  & 95.28 & 94.87   & 95.24 & 93.78 & 92.20\\ \dbottomrule
\end{tabular}%
}
\caption{Results demonstrating the transferability of robustness to different attack methods. The evaluated model was trained on PGD $L_\infty$.}
\label{tab:transfer_robustness}
\vspace{-5mm}
\end{table}

\section{Conclusion}
We proved and demonstrated that training a robust malware detector on feature-space adversarial examples inherently generates robustness against problem-space malware samples. Subsequently, we proposed a Bayesian adversarial learning objective in the feature space to realize a robust malware detector in the problem space. Additionally, we explain the improved performance by proving that our proposed method bounds the difference between adversarial risk versus empirical risk to improve robustness and show the benefits of a BNN as a defense method (see \textbf{Appendix}).
Our empirical results, including a production scale dataset, demonstrates new state-of-the-art \textit{performance} and \textit{robustness} benchmarks.

\section{Acknowledgments}

This research was supported by the Next Generation Technologies Fund from the Defence Science and Technology Group, Australia. 
We also thank Dr. Sharif Abuadbba for supporting us in collecting malware samples for the project.

\bibliography{main}


\newpage
\appendix

\onecolumn
\section{Appendix}
\subsection{Proof of Lemma 1}
\label{appd:proof_theorem1}

We consider the problem-space example $\bz \in \Z$ with the feature in the feature space $\bx \in \X$. Initially, we assume that there is no solution to the feature-space attack ($\bx'$) under constraints $\Upsilon$ given an existing problem-space adversarial attack ($\bz'$) under constraints $\Omega$. Specifically, we assume that there is no $\bdelta$ that satisfies $p(y\mid\bx + \bdelta, \btheta) < 0.5$, $\bdelta \models \Upsilon$. However, because there exists an adversarial attack in the problem-space,
\begin{equation}
    \exists \bz': \bz' \models \Omega, \quad p\big(y \mid \Phi(\bz'), \btheta\big) = p(y\mid\Phi(\bT(\bz)), \btheta),  \quad p(y\mid\Phi(\bT(\bz)), \btheta) < 0.5
\end{equation}

\noindent Thus,
\begin{equation}
    \label{NeurIPS_eq:T}
    \exists \bT^* : p(y\mid\Phi(\bz'),\btheta) = p(y\mid\Phi(\bT^*(\bz)),\btheta), p(y\mid\Phi(\bT^*(\bz)),\btheta) < 0.5, \quad \bz, \bz' \models \Omega
\end{equation}
For any transformation $\bT$ in the input space, there exists a value $\bdelta^*$ for which we have: 
\begin{equation}
    \exists \bT^* : p(y\mid\Phi(\bT^*(\bz)),\btheta) = p(y\mid\bx + \bdelta^*,\btheta), \quad p(y\mid\bx + \bdelta^*,\btheta)  < 0.5, \quad \bz, \bz' \models \Omega
\end{equation}

We recall that feature-space constraints are determined by problem-space constraints, that is, the search space allowed by $\Omega$ is stricter or equal to that allowed by $\Upsilon$. Thus, $\bdelta^*$ must be a solution to feature-space constraints $\Upsilon$.
However, this is not possible because we begin with the assumption that there is no $\bdelta$ that satisfies $f(\bx + \bdelta) < 0.5, \bdelta \models \Upsilon$. This contradiction proves that:
\begin{align*}
&\exists \bz': \bz' \models \Omega, \quad p(y\mid\Phi(\bz'),\btheta) = p(y\mid\Phi(\bT(\bz)),\btheta), \quad p(y\mid\Phi(\bT(\bz)),\btheta) < 0.5 \\
&\Rightarrow \exists \bx'=  \bx + \bdelta: \bx' \models \Upsilon, p(y\mid\bx',\btheta) < 0.5
\end{align*}

\subsection{Proof of the Objective (Proposition 1)}
\label{sec:appd_proof}
We have
\begin{align*}\left|R_{\text{adv}}-R\right| & =\left|\mathbb{E}_{(\bx,y)\sim\D}\Bigg[\mathbb{E}_{\btheta}\Bigg[\underset{}{\sup}\,\,\mathbb{E}_{y_{1}\sim p(y|\bxa)}\left[\II\left(y_{1}\neq y\right)\right]-\mathbb{E}_{y_{2}\sim p(y|\bx)}\left[\II\left(y_{2}\neq y\right)\right]\Bigg]\Bigg]\right|\,,\\
 & =\left|\mathbb{E}_{(\bx,y)\sim\D}\Bigg[\mathbb{E}_{\btheta}\Bigg[\sup\,\,\mathbb{E}_{y_{1}\sim p(y|\bxa),y_{2}\sim p(y|\bx)}\left[\II\left(y_{1}\neq y\right)-\II\left(y_{2}\neq y\right)\right]\Bigg]\Bigg]\right|\,,\\
 & \quad\left.\leq\mathbb{E}_{(\bx,y)\sim\D}\Bigg[\mathbb{E}_{\btheta}\Bigg[\sup\,\,\mathbb{E}_{y_{1}\sim p(y|\bxa),y_{2}\sim p(y|\bx)}\left[|\II\left(y_{1}\neq y\right)-\II\left(y_{2}\neq y\right)|\right]\Bigg]\right]\,,\\
 & \quad\leq\mathbb{E}_{(\bx,y)\sim\D}\Bigg[\mathbb{E}_{\btheta}\Bigg[\sup\,\,\mathbb{E}_{y_{1}\sim p(y|\bxa),y_{2}\sim p(y|\bx)}\left[\II\left(y_{1}\neq y_{2}\right)\right]\Bigg]\Bigg]\,.
 \end{align*}
 where we can upper bound the expected misclassification to have:

 \begin{align*}
 &\mathbb{E}_{(\bx,y)\sim\D}\Bigg[\mathbb{E}_{\btheta}\bigg[1-\sum_{c=1}^{K}p(y=c\mid\bx,\btheta)p(y=c\mid\bxa,\btheta)\bigg]\Bigg]\,.
 \end{align*}
 Subsequently, we use Jensen's inequality and the fact that $\bx=\exp(\log(\bx))$ to have:
 \begin{align*}
 &\mathbb{E}_{(\bx,y)\sim\D}\Bigg[\mathbb{E}_{\btheta}\bigg[1-\exp(\underbrace{\log(\sum_{c=1}^{K}p(y=c\mid\bx,\btheta)p(y=c\mid\bxa,\btheta))}_{\geq \sum_{c}^{K}p(y=c\mid\bx,\btheta)\log(p(y=c\mid\bxa,\btheta)})\bigg]\Bigg]\,,
 \end{align*}
 and since $1-\exp(z)$ is monotonically decreasing, we have
 \begin{align*}
 &\mathbb{E}_{(\bx,y)\sim\D}\Bigg[\mathbb{E}_{\btheta}\bigg[1-\exp({\log(\sum_{c=1}^{K}p(y=c\mid\bx,\btheta)p(y=c\mid\bxa,\btheta))})\bigg]\Bigg]\\
 & \leq \mathbb{E}_{(\bx,y)\sim\D}\Bigg[\mathbb{E}_{\btheta}\bigg[1-\exp\big(\sum_{c}^{K}p(y=c\mid\bx,\btheta)\log(p(y=c\mid\bxa,\btheta))\big)\bigg]\Bigg]\\
 & \quad\quad= 1-\mathbb{E}_{(\bx,y)\sim\D}\Bigg[\mathbb{E}_{\btheta}\bigg[\exp\big(\sum_{c}^{K}p(y=c\mid\bx,\btheta)\log(p(y=c\mid\bxa,\btheta))\big)\bigg]\Bigg]\,.\\
 & \qquad\quad
\end{align*}

Thus we have the following bound:
\begin{align}
\label{eq:bound}
\left|R_{\text{adv}}-R\right|\leq1-\mathbb{E}_{(\bx,y)\sim\D}\Bigg[\exp\Bigg(\mathbb{E}_{\btheta}\bigg[\underbrace{\sum_{c}^{K}p(y=c\mid\bx,\btheta)\log(p(y=c\mid\bxa,\btheta))}_{r_{\btheta}(\bx,\bxa,y)}\bigg]\Bigg)\Bigg]\,.
\end{align}
This result demonstrates that the difference between the risks is bounded by the negative cross entropy of the predictions. While informative, this bound expresses the  relation between the predictions only and not how the model perform on each set (\ie given dataset versus its corresponding adversarial).

Then the difference between the empirical risk and the adversarial risk is minimized when the upper bound is minimized. Hence, the main learning objective is to:
\begin{itemize}[label={}]
    \item Maximize $\mathbb{E}_{\btheta}[\sum_{c}^{K}p(y=c\mid\bx,\btheta)\log(p(y=c\mid\bxa,\btheta))]$ or minimize $\mathbb{E}_{\btheta}[\text{KL}(p(y=c\mid\bx,\btheta)\|p(y=c\mid\bxa,\btheta))$: this corresponds to matching the prediction from the adversarial data to that of the observations. Since $(\bx,y)$ is given in training, for minimizing this KL-divergence we simply minimize the entropy of the adversarial examples instead;
\end{itemize}
Notably, since we know $1-\exp(-z)\leq z$, to avoid computational instabilities and gradient saturation, we consider minimizing the upper bound without the exponential function in our implementation.

\subsection{Datasets}
\label{appd:dataset}
Below is the detailed information regarding the datasets used in the paper: 

\begin{itemize}[leftmargin=*]
    \item The \noindent\textbf{\textsf{EMBER}} dataset contains portable executable files (PE files) scanned by VirusTotal on or before 2018. It includes 600,000 training samples and 200,000 testing samples, equally distributed between benign programs and malicious programs, where the malware is labeled by the malware family using AVClass~\citep{sebastian2016avclass}. This dataset also includes the vectorized features which encode a great amount of information regarding the PE files, such as general file information, import/export functions, header information, and string information. This EMBER dataset (2018) was designed to be \textit{more challenging} for ML-based classifiers compared with the older version --- the EMBER2017 dataset. Thus, using this dataset is of interest to evaluate our ML-based malware classifier.
    \item The \noindent\textbf{\textsf{SOREL-20M}} dataset is a recent industrial-scale dataset from Sophos AI~\citep{harang2020sorel}. It contains 20 million samples with their pre-extracted features and metadata and high-quality labels. Particularly, the dataset includes 12,699,013 training samples, 2,495,822 validation samples, and 4,195,042 testing samples. The dataset also provides 10 million disarmed malware samples for feature exploration. With a significant amount of high-quality samples, this dataset aims to provide the new default benchmark for malware detection.
\end{itemize}
 
Note that both datasets follow a strict temporal split policy where test samples were observed strictly later than training samples. 

\subsection{Feature Computations}

The paper explores the use of pre-extracted features rather than raw bytes. While the process of manually extracting these features may require significant effort from domain experts, recent developments in the field~\cite{anderson2018ember, harang2020sorel} have automated this process and developed techniques to extract raw bytes information, which now takes only an average of 160 ms on our hardware using an NVIDIA RTX A6000 GPU to convert binary malware to its corresponding vectorized feature. This advancement makes it possible to employ feature-based methods such as adversarial learning in large-scale datasets.

\subsection{Hyper-Parameters and Network Architecture}
\label{sec:appd_hyper}
\label{appd:network_architecture}

In this section, we provide detailed information about the hyper-parameters and network architecture utilized in our experiments. Table~\ref{table:hyperparameters} presents the specific parameters employed in PGD to generate adversarial examples, such as $\alpha$ and $\varepsilon_{\max}$, as well as the weight $\gamma$ used to regulate the repulsive force in SVGD.

Regarding the network architecture, we employ the feed-forward neural network (FFNN) architecture provided in SOREL-20M~\citep{harang2020sorel} as the baseline for our Bayesian neural network versions. This architecture consists of four fully-connected (FC) layers. It is worth noting that there are no baseline deep neural networks available in the EMBER dataset~\citep{anderson2018ember}. Therefore, we utilize the same baseline network architecture presented in Table~\ref{tab:sorel_arch} for the EMBER dataset.

\begin{table}[h!]
\centering
\small
\begin{adjustbox}{width=0.8\linewidth, center}
\begin{tabular}{ccc} 
\dtoprule
\textit{Name} & \textit{Value} &  \textit{Notes} \\
\midrule
$T$ & 20 & \#PGD iterations \\ \midrule
$\varepsilon_{\max}$ & 0.03 & Max $l_\infty$-norm in adversarial training \\ \midrule
$\alpha$ & 0.02 & Step size for each PGD iteration \\ \midrule
$\gamma$ & SOREL-20M:0.01, EMBER:0.05 & Weight to control the repulsive force \\ \midrule
$n$ & SOREL-20M:5, EMBER:20 & \makecell{\#Parameter particles \\ \#Forward passes when doing ensemble inference \\ \# Expectation over Transformation} \\ \midrule

\end{tabular}
\end{adjustbox}
\caption{{Hyper-parameter settings in our experiments}}
\label{table:hyperparameters}
\end{table}

\begin{table}[h!]
\centering
\begin{adjustbox}{width=.5\linewidth, center}

\begin{tabular}{cccc}
\dtoprule
Layer Type & \# of Channels  & Drop out & Activation \\ \midrule
FC       & 512      & 0.05                & ELU       \\
FC       & 512      & 0.05                & ELU       \\
FC       & 128      & 0.05                & ELU       \\
FC       & 1       & -                & Sigmoid       \\ \dbottomrule
\end{tabular}
\end{adjustbox}
\caption{Model Architecture for SOREL-20M and EMBER}
\label{tab:sorel_arch}
\vspace{2mm}
\end{table}

\subsection{Metrics}
\label{appd:metrics}
Here, we introduce the metrics commonly used to evaluate malware detectors in the literature in order to quantify experimental results. 

\begin{itemize}
  \item \emph{True Positive (TP)} indicates the number of predictions where the network correctly predicts the malware samples as malware.
  \item \emph{True Negative (TN)} indicates the number of predictions where the network correctly predicts the benign ware as benign ware.
  \item \emph{False Positive (FP)} indicates the number of predictions where the network incorrectly predicts the benign ware samples as malware.
  \item \emph{False Negative (FN)} indicates the number of predictions where the network incorrectly predicts the malware samples as benign ware.
\end{itemize}

\noindent\textbf{Accuracy.} We compute the overall accuracy of the model, denoted by $\text{Acc}$, to quantify the proportion of samples that were correctly classified by the classifier:
\begin{equation*}
  \mathrm{Acc} =\frac{\mathrm{TP}+\mathrm{TN}}{\mathrm{TP}+\mathrm{TN}+\mathrm{FP}+\mathrm{FN}}
\end{equation*}

\heading{Robustness}. This metric, denoted by $Robustness$, quantifies the accuracy of the model under the threat of adversarial attacks according to the following equation:

\begin{equation*}
  \mathrm{Robustness} = \mathrm{Acc}(\bxa), \quad \bxa \sim \Da
\end{equation*}

\noindent\textbf{Attack success rate.} This metric, denoted by $\text{ASR}$, quantifies the proportion of samples that were incorrectly classified by the classifier according to the following equation:
\begin{equation*}
  \mathrm{ASR}=\frac{\mathrm{FP}+\mathrm{FN}}{\mathrm{TP}+\mathrm{TN}+\mathrm{FP}+\mathrm{FN}}
\end{equation*}

\noindent\textbf{True positive rate (TPR).} This metric, also known as \emph{Recall or Sensitivity}, quantifies the fraction of all malware samples that the classifier correctly predicted as malware according to the following equation:

\begin{equation*}
  \mathrm{TPR}=\frac{\mathrm{TP}}{(\mathrm{TP}+\mathrm{FN})}
\end{equation*}

\noindent\textbf{False positive rate (FPR).} This metric, equivalent to \emph{(1-Specificity)}, where \emph{Specificity} is the TPR, quantifies the fraction of all benign ware samples that the classifier correctly predicted as benign ware according to the following equation:


\begin{equation*}
  \label{eq:FPR}
  \mathrm{FPR}=\frac{\mathrm{FP}}{(\mathrm{FP}+\mathrm{TN})}
\end{equation*}

\heading{Receiver Operating Characteristics (ROC) and Area Under the Curve (AUC).} All possible combinations of TPR and FPR compose a Receiver Operating Characteristics space, denoted by ROC, with each point in ROC space determined by a pair (TPR, FPR), showing the trade-off between \emph{ Sensitivity } and \emph{Specificity}. Based on this ROC curve, we can calculate Area Under the Curve (AUC) to evaluate the performance of the detector.

\newpage

\subsection{Adversarial Bayesian Learning Algorithm for \ours}
\label{appd:algorithm}

\begin{algorithm}[h!] %
\caption{Adversarial Bayesian Learning with SVGD}  
\label{alg:alg1}
\begin{algorithmic}[1]
\STATE {\bf Input:} A set of initial parameter particles $\{\btheta_i^0\}_{i=1}^n$, observation feature-space data $\D$. 
\STATE {\bf Output:} A set of parameter particles $\bTheta:=\{\btheta_i\}_{i=1}^n$ that approximates the true posterior distribution $p(\btheta\mid\Da)$ 
\FOR{$(\bx,y)\sim p(\D)$}
\STATE $\bxa\gets\bx$
\FOR{$t=1\to T$} 
\STATE $\begin{aligned}
\bxa=\Pi_{\varepsilon_{\max}}\left\{\right.\bxa&+\alpha \cdot \operatorname{sign}\left(\right. 
&\mathbb{E}_{\btheta}\left[\nabla_{\bx} \ell\left(f\left(\bxa ; \btheta_j\right), y\right)\right]\left.\right)\left.\right\}
\end{aligned}$
\COMMENT {Generate feature-space adversarial samples (Eq.~\eqref{eq:a_pgd})}
\ENDFOR
\FOR{$i=1\to n$}
\STATE $\btheta_i  \gets   \btheta_i - \epsilon_i \hatff{}^*(\btheta_i, \btheta_j) $ 
\text{\quad with}
$\hatff{}^*(\btheta_i,\btheta_j) = \sum_{j=1}^n\big[  k(\btheta_j, \btheta_i)  \nabla_{\btheta_j} \ell(\bTheta) - \frac{\gamma}{n}\nabla_{\btheta_j} k(\btheta_j, \btheta_i)\big]$
\STATE $\epsilon_i$ is the step size at the current iteration, $k(\btheta,  \btheta')$ is a positive definite kernel that specifies the similarity between $\btheta$ and $\btheta'$, $\ell$ is binary cross entropy loss, $\gamma$ is the weight to control the \textit{repulsive force} that enforces the diversity among parameter particles.
\ENDFOR
\ENDFOR
\end{algorithmic}
\end{algorithm}

\subsection{The Impact of the Number of Parameter Particles for Capturing the Posterior Distribution}

Here, we investigate the contribution of the number of parameter particles to the robustness of the networks. As shown in Table~\ref{tab:ablative_particle}, the robustness of the BNNs is improved when we increase the number of particles from 5 to 10. This is intuitive because increasing the number of parameter particles will help to capture the multi-modal posterior better. Thus, enlarging the number of parameter particles may further improve the network's robustness.

\begin{table}[h!]
\centering
\vspace{-2mm}
\resizebox{.6\linewidth}{!}{%
\begin{tabular}{ccccccc}
\dtoprule
 \multirow{2}{*}{\ours Networks} & \multicolumn{6}{c}{Attack budget} \\ \cmidrule{2-7}
 & \textit{0}      & \textit{0.03}  & \textit{0.05}    & \textit{0.1}   & \textit{0.2}   & \textit{0.3}   \\ \dtoprule
5 particles              & 95.52  & 94.20 & 90.53   & 62.86 & 25.42 & 23.10 \\ \midrule
10 particles                & \textbf{96.29}  & \textbf{94.97} & \textbf{92.19}   & \textbf{69.96} & \textbf{35.20} & \textbf{30.79} \\ \dbottomrule
\end{tabular}%
}
\caption{Assessing the contribution of number of parameter particles to the robustness of the networks on the SOREL-20M dataset.}
\label{tab:ablative_particle}
\vspace{-2mm}
\end{table}

\subsection{Additional Experimental Results}
\label{appd:results}
\vspace{2mm}

In this section, we provide additional results to support the statements in the main paper further.

\vspace{1mm}
\noindent\textbf{Performance Comparison with MalConv~\citep{raff2018malware}.~} We remark that training on raw byte sequences (MalConv) is computationally expensive; hence, there is no available MalConv on the large production scale dataset (SOREL-20M) for comparison. In this experiment, we want to highlight that the FFNN and BNN networks that were used in our experiments, whilst being more efficient in scaling up to a large production scale, outperform MalConv. In addition, we also notice that MalConv did not perform well on the adversarial malware as shown in the main paper under Section \textbf{Robustness (against Problem-Space Adversarial Malware)}.

We provide, in more detail, the performance comparison of networks on the EMBER test set under no attacks in Table~\ref{tab:noattacks}. Here, we compared FFNN and BNN networks trained on pre-extracted features with the popular ML-based malware detector built on problem-space samples (MalConv). In particular, we show that a Bayesian neural network built on the feature-space samples achieves better performance compared with MalConv and its recently upgraded version in AAAI'21 (MalConv w/ GCG)~\citep{raff2020classifying}. 

\begin{table}[h!]
\centering
\begin{tabular}{cccc}
\dtoprule
Dataset & Network &
  Acc &
  AUC \\ \midrule
 
 \multirow{4}{*}{EMBER} & 
MalConv~\citep{raff2018malware} &
  90.88 &
  97.05 \\ \cmidrule{2-4}
& MalConv w/ GCG~\citep{raff2020classifying} &
  93.29 &
  98.04 \\ \cmidrule{2-4}
&  FFNN &
  94.11 &
  98.47 \\ \cmidrule{2-4}
& BNN &

  \textbf{94.50} &
  \textbf{98.55} \\ \dbottomrule

\end{tabular}
\caption{Comparing performance of BNN and FFNN with MalConv}
\label{tab:noattacks}
\end{table}

\vspace{1mm}
\noindent\textbf{Comparison with~\citep{advbnn} Learning Objective for Adversarially Training a BNN.} We also compare ours with a previous method for adversarially training a Bayesian neural network~\cite{advbnn} and report the results in Table~\ref{tab:compare_stochastic} where we show the significantly better performance achieved with our training method.

\begin{table}[h]
\centering
\resizebox{.7\textwidth}{!}{%
\begin{tabular}{@{}cccccccc@{}}
\toprule
Dataset                    & Networks              & 0              & 0.03           & 0.05           & 0.1            & 0.2            & 0.3            \\ \midrule
\multirow{2}{*}{SOREL-20M} & Adv-BNN (~\citet{advbnn}) & 94.56          & 92.97          & 89.31          & 56.69          & 17.39          & 14.95          \\ \cmidrule(l){2-8} 
                           & \ours (Ours)       & \textbf{95.52} & \textbf{94.20} & \textbf{90.53} & \textbf{62.86} & \textbf{25.42} & \textbf{23.10} \\ \bottomrule
\end{tabular}%
}
\caption{Comparing the performance of Adversarially trained Bayesian Neural Networks}
\label{tab:compare_stochastic}
\end{table}

\begin{figure}[h!]
    \centering
    \includegraphics[width=.5\linewidth]{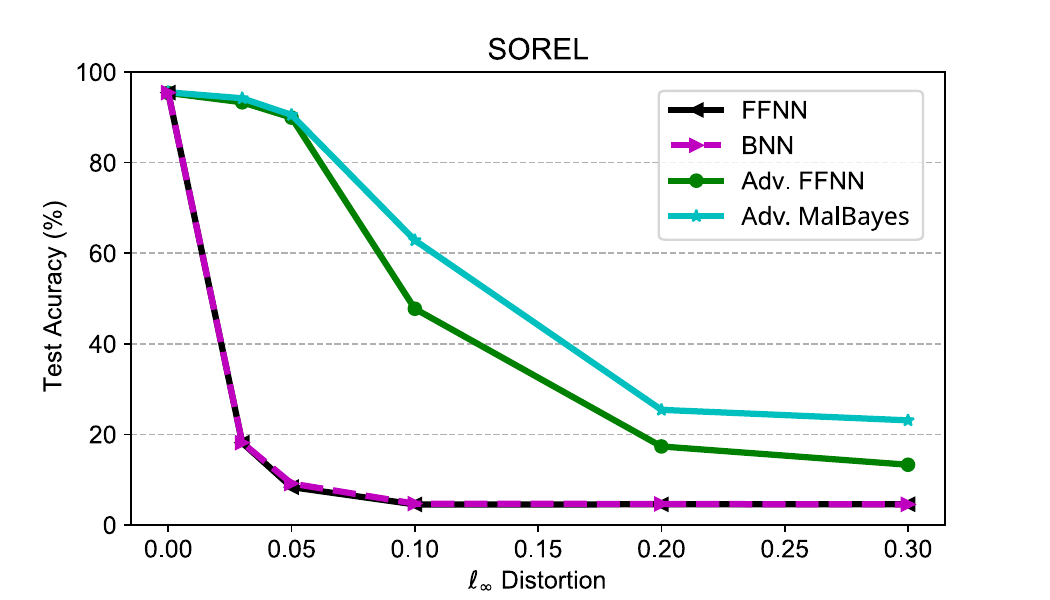}
    \caption{Comparing performance of different training method under adversarial attacks generated in the feature space with increasing perturbation budgets. Adversarially trained networks are trained with the budget of $\epsilon=0.03$.} 
    \label{fig:attack_results}
\end{figure}

\vspace{1mm}
\noindent\textbf{Comparing Performance of Training Methods under Adversarial Attacks.~}We also conduct experiments to show the performance of FFNN, BNN, Adv-FFNN, and \ours under feature-space adversarial attacks with increasing attack budgets. The results in Figure~\ref{fig:attack_results} shows that, as expected, the adversarial training methods yield improved robustness. 

\begin{figure}[h!]
    \centering
    \includegraphics[width=.6\linewidth]{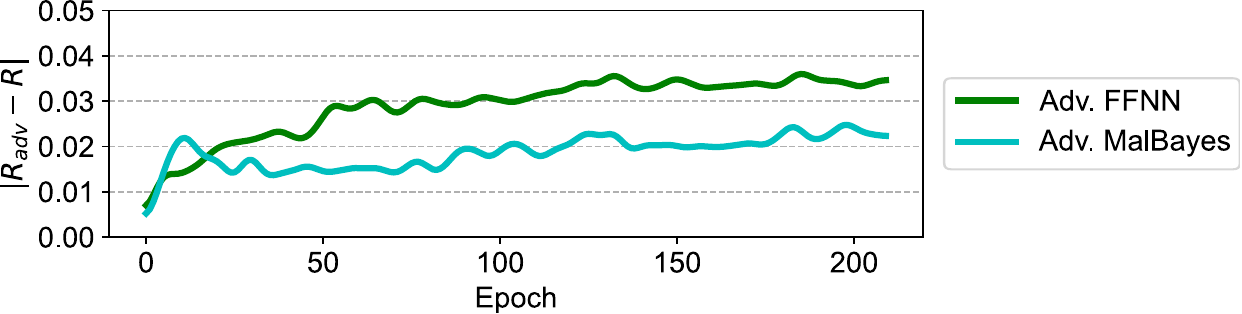}
    \caption{The gap between conventional empirical risk and adversarial risk $|R_{adv} - R|$} evaluated on the test set of \textsf{EMBER}.
    \label{fig:R_gap}
\end{figure}

\vspace{1mm}
\noindent\textbf{Empirically Demonstrating the Impact of BNN formulation to Bound the Difference Between Adversarial Risk and Empirical Risk.~}
We conduct the empirical experiment to show the difference between adversarial risk and empirical risk, illustrated in Figure~\ref{fig:R_gap}. Here, we measure the empirical risk and its adversarial counterpart on the test set of EMBER dataset. We can observe the benefit of hardening a BNN with adversarial examples, which helps to tighten the gap between the adversarial and empirical risks and improve robustness compared with the non-Bayesian method (Adv-FFNN) demonstrating the theoretical bound mentioned in Eq.~\eqref{eq:bound}. 

\vspace{3mm}
\noindent\textbf{Generalization to Another Software Domain (Android)}
To demonstrate the generalizability of our approach, we assess the robustness of our proposed method in another software domain, specifically Android malware. We adopt the same methodology and train the corresponding FFNN, BNN, Adv-FFNN, and Adv-MalBayes networks on a widely used Android DREBIN dataset~\cite{arp2014drebin}. Given that the feature space of the DREBIN dataset is binary, we utilize the FGSM attack to conduct adversarial learning. The results presented in Table~\ref{tab:robustness_android} yield a similar observation to that obtained in the Windows PE domain. Our proposed Adv-MalBayes outperforms Adv-FFNN, while other models, such as FFNN, are entirely defeated by adversarial malware examples. Thus, our approach demonstrates generalization across various software domains.

\begin{table}[h]
\centering
\resizebox{.6\linewidth}{!}{%
\begin{tabular}{cccccccc}
\dtoprule 
\multirow{2}{*}{Dataset} & \multirow{2}{*}{Networks} & \multicolumn{6}{c}{Attack budget} \\ \cmidrule{3-8}
 &  & \textit{0}      & \textit{0.01}  & \textit{0.03}    & \textit{0.1}   & \textit{0.2}   & \textit{0.3}   \\ \dtoprule
\multirow{2}{*}{DREBIN} & Adv-FFNN               & 92.26  & 88.63 & 86.42 & 76.19 & 63.11 & 57.38  \\ \cmidrule{2-8}
& \ours                & \textbf{94.62}  & \textbf{92.34} & \textbf{89.06}   & \textbf{85.27} & \textbf{82.2} & \textbf{80.34} \\ \dbottomrule
\end{tabular}%
}
\caption{Robustness of networks against adversarial Android malware.}
\label{tab:robustness_android}
\end{table}

\vspace{3mm}

\noindent\textbf{Transferability of Robustneess Against PGD $L_2$ Attack.}
In Section \textbf{Transferability of Robustness} in the main paper, we have evaluated the  robustness of our proposed method to FGSM attack~\citep{goodfellowExplainingHarnessingAdversarial2015}  because most existing attacks~\citep{suciu2019exploring} adopt from it. This section will further evaluate the transferability of our method on PGD $L_2$ attacks. Using the based model Adv-MalBayes trained on PGD $L_\infty$, we conduct PGD $L_2$ attacks on 10,000 random malware samples, and Table~\ref{tab:transfer_robustness_1} shows that Adv-MalBayes is highly robust and still maintains its robustness against PGD $L_2$ attacks in the most challenging attack budget we consider.

\begin{table}[h]
\vspace{-2mm}
\centering
\resizebox{.5\linewidth}{!}{%
\begin{tabular}{ccccccc}
\dtoprule
 \multirow{2}{*}{\ours Networks} & \multicolumn{6}{c}{{Attack budget}} \\ \cmidrule{2-7}
 & \textit{0}      & \textit{0.03}  & \textit{0.05}    & \textit{0.1}   & \textit{0.2}   & \textit{0.3}   \\ \dtoprule
PGD $L_\infty$              & 96.29  & 94.97 & 92.19   & 69.96 & 35.20 & 30.79 \\ \midrule

PGD $L_2$              & -  & 95.13 & 95.08   & 95.03 & 94.87 & 94.85\\ \dbottomrule
\end{tabular}%
}
\caption{Results demonstrating the transferability of robustness to the PGD $L_2$ attack method. The evaluated model was trained on PGD $L_\infty$.}
\label{tab:transfer_robustness_1}
\end{table}

\noindent\textbf{Robustness Against Partial DOS Header Attack.~}
In this section, we compare the performance of our proposed \textit{Adv-MalBayes} approach against another problem-space attack technique, specifically Partial DOS Header Manipulation~\cite{demetrio2019explaining}. We utilize the implementation provided in \textit{SecML-malware}~\cite{demetrio2021secml} with a configuration of 200 iterations. To demonstrate the robustness of our approach against this attack, we evaluate it on 1,000 randomly selected real-world malware samples. The results indicate that our proposed method remains completely robust (i.e., unaffected by the problem-space attack), while the robustness of MalConv degrades to only 65\%. This finding further confirms the effectiveness of our approach in dealing with actual, functional malicious software.



\end{document}